# INTERACTIONS OF YOUNG BINARIES WITH DISKS

STEPHEN H. LUBOW
*Space Telescope Science Institute*

and

PAWEL ARTYMOWICZ
*Stockholm Observatory*

The environment of a binary star system may contain two circumstellar disks, one orbiting each of the stars, and a circumbinary disk orbiting about the entire binary. The disk structure and evolution are modified by the presence of the binary. Resonances emit waves and open disk gaps. The binary's total mass and mass ratio as well as orbital elements can be modified by the disks. Signatures of these interactions provide observational tests of the dynamical models. The interaction of young planets with protoplanetary disks circularizes the orbits of Jupiter-mass planets and may produce much more massive extrasolar planets on eccentric orbits.

## I. INTRODUCTION

The study of the interaction of young binary stars with disks is motivated by two factors: the high frequency of binarity in young stars and the high frequency of disks around young stars. Most stars, including young stars, are found in binary star systems. Although the binary formation process is as yet not well understood (see Bodenheimer et al. 1993; the chapter by Bodenheimer et al., this volume), binarity appears to be established among the youngest observed stars (Ghez et al. 1993; Mathieu 1994).

General arguments suggest that disks are an inevitable consequence of the star formation process. The standard picture of single-star formation involves the collapse of a rotating cloud (Shu et al. 1987). A disk is a natural product of the collapse, because centrifugal forces prevent much of the cloud material from falling directly onto the central star. Instead, the material settles onto a centrifugally supported disk, from which accretion can occur onto the central star (Lynden-Bell and Pringle 1974).

Considerable observational evidence now exists for the presence of disks around young stars. Observationally determined disk frequencies in star-forming regions exceed 50% (see review by Sargent and Beckwith 1994). The images of disks taken with the *Hubble Space Telescope* (HST) provide overwhelming evidence for the existence of disks (e.g., McCaughrean and O'Dell 1996; Burrows et al. 1996). Resolved circumbinary disks are now directly observed (e.g., Dutrey et al. 1994; Roddier et al. 1996).

The scale of a typical binary star separation is about 30–50 AU for field stars (Duquennoy and Mayor 1991), which is less than the typical T Tauri disk size of a few hundred AU. Consequently, one expects that binary star systems will usually interact strongly with disks (Beckwith et al. 1990). Except for the shortest-period systems (periods less than about 8 days), binaries are eccentric with typical eccentricity of about 0.3 (see Mathieu 1994).

These interactions give rise to several phenomena important in evolutionary and observational contexts. In this review we emphasize the recent theoretical results obtained after the publication of a previous extensive review by Lin and Papaloizou (1993).

## II. GENERAL DISK PROPERTIES

### A. Disk Configuration

There are two types of disks in a binary environment: circumstellar (CS) disks, which surround only one star, and circumbinary (CB) disks, which surround the entire binary. Two CS disks, one around each star, and one CB disk can be present in a binary. The CS and CB disks are separated by a tidally produced gap (e.g., Lin and Papaloizou 1993; Artymowicz and Lubow 1994).

The exact size of the gap depends on several factors, including the binary eccentricity, disk turbulent viscosity, and sound speed, as will be described later. Subject to certain assumptions, the inner edge of the CB disk is typically $1.8a$ to $3a$, for binary semimajor axis $a$. The circumprimary and circumsecondary disk outer edges have typical radii of $0.35a$ to $0.5a$ and $0.2a$ to $0.3a$, respectively, for a binary with a mass ratio of 3:1.

Support for this picture comes from studies of millimeter (and submillimeter) emission of young binary systems, which is characteristic of material at several tens of AU from a central star. For binaries with (projected) separations of that scale, there is a systematic deficit of emission (Beckwith et al. 1990; Jensen and Mathieu 1997). On the other hand, near-IR emission, characteristic of regions close to each star, often does not show a corresponding deficit. This result is suggestive of gaps produced on the size scale of the binary.

The most direct evidence for gaps in disks comes from observations of binaries GG Tauri and UY Aurigae. In each case of these two systems, both millimeter interferometry and IR imaging clearly show the existence of a CB disk and a depletion of material near the central regions where the binary is located (Duvert et al. 1998; Close et al. 1998). For both binaries, there are observational signatures of circumstellar disks and accretion onto the stars (Hartigan et al. 1995; Hartmann et al. 1998).

An important related issue is whether these gaps are completely clear of material or whether material might flow through the gap and influence the evolution of the binary (section III.F)

In most cases, this review will assume that the disks are coplanar with the binary orbit. Although this assumption is a natural one, given the geometry of a simple-minded cloud collapse model, it need not actually be realized. Recent theoretical and observational results suggest that noncoplanarity sometimes occurs (cf. section III.E)

## B. Evolutionary State

In the earliest phases of stellar evolution (Classes 0 and I of Lada and Shu 1990), a young protobinary might be expected to be embedded in its parent cloud. Cloud material accretes onto the binary and the disks. The infalling material affects the disk structure and accretion properties (e.g., Cassen and Moosman 1981).

Modeling binary evolution in this early phase has emphasized the evolution of the binary orbit and mass ratio, assuming the binary orbit remains circular at all times. The main result is that the binary mass ratio responds to the specific angular momentum (angular momentum per unit mass) of the accreting cloud material, relative to the specific angular momentum of the binary (Artymowicz 1983; Bate and Bonnell 1997). If the material falls in with low specific angular momentum (lower than approximately the binary specific angular momentum), then it is accreted preferentially onto the primary star. The reason is that the matter falls inward toward the deeper potential well. With high-angular-momentum accretion, the material preferentially accretes onto the secondary star. The high angular momentum of the material holds it at larger radii, where it can more easily encounter the secondary (for a physical argument see section III).

The semimajor axis evolution of a circular-orbit binary is dominated by gravitational torques of the accreting material and the advection of angular momentum carried by the accreting material. Gravitational torques cause the orbit to shrink. Generally, it is found that the advection of angular momentum dominates over gravitational torques. So the overall sense of the orbit evolution is that the orbit shrinks when material with low specific angular momentum is accreted, and the orbit expands when material with high specific angular momentum is accreted.

Figure 1 displays a set of smoothed particle hydrodynamics (SPH) results of the accretion flow for different values of specific angular momentum in the inflowing material. The embedding cloud that supplies material is likely to have its specific angular momentum increase with radius in order to be stable dynamically. Consequently, the specific angular momentum of material supplied to the binary will likely increase in time, as material collapses onto the binary from greater and greater distances. The frames in Fig. 1 can then be considered to represent a sequence in time.

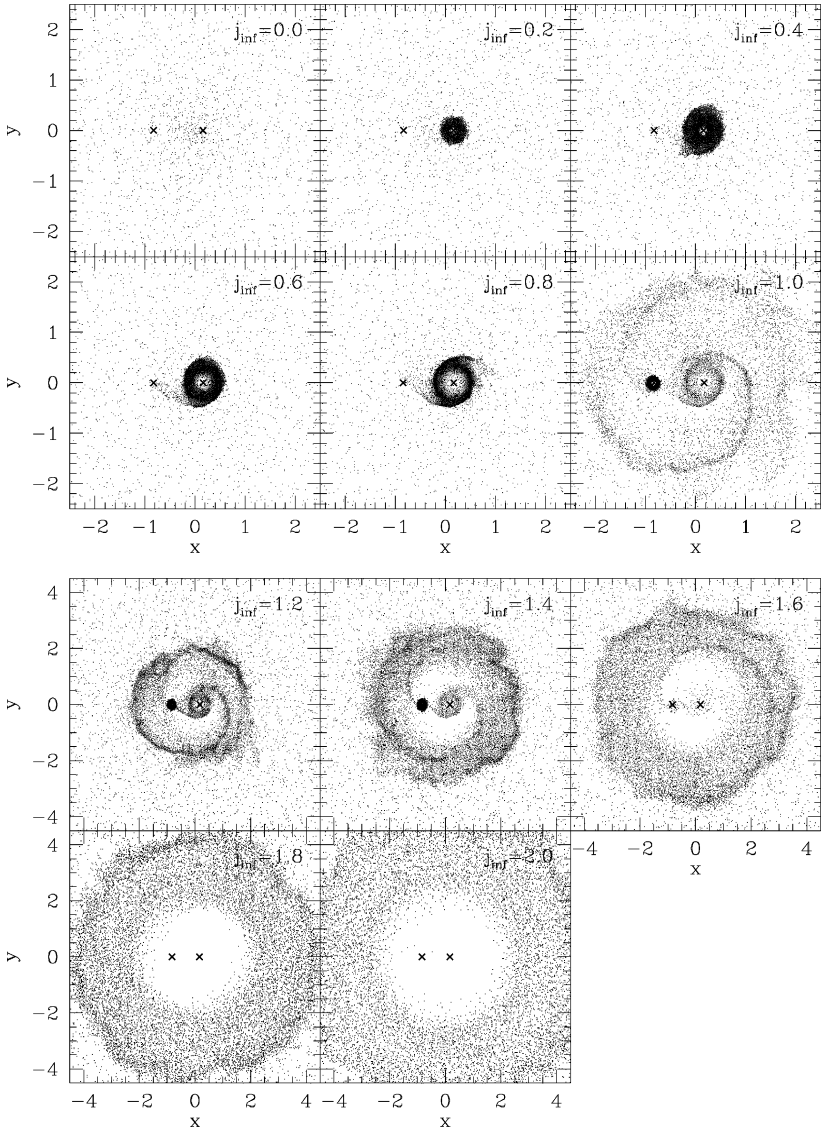

Figure 1. SPH simulations of the accretion onto the protobinary star of mass ratio 0.2 for different values of specific angular momentum in the inflowing material, $j_{\rm inf}$, determining the partitioning of the flow between the circumstellar disks (and stars). (Taken from Bate and Bonnell 1997.)

Starting with the lowest-angular-momentum inflow (earliest times), the sequence reveals that a Bondi-Hoyle (Bondi and Hoyle 1944) accretion column develops about each star, followed by the development of disks around first the primary star and then the secondary star. For inflow with somewhat higher specific angular momentum, a circumbinary disk forms, but the angular momentum is still low enough to drive an inflow into the binary. Ongoing observational work on embedded sources will provide more constraints on this early evolutionary phase (e.g., Looney et al. 1997).

For sufficiently high specific angular momentum, the disk settles into Keplerian orbits that surround the entire binary. Subsequent evolution occurs on the viscous timescale of the disk.

At later times, the binary consists of T Tauri (or Herbig Ae/Be) stars and is no longer surrounded by the cloud (Class II). Remnant disks are present in the system. The system may have a circumbinary disk, as seen in the last frame of Fig 1. In addition, remnant circumstellar disks may be present, which may have been formed during a phase of accretion of lower specific angular momentum, as seen in the middle frames of Fig. 1. To date, more attention has been given to the properties of binary systems at this state, because such systems are more easily observed. The remainder of this review will concentrate on such systems.

### C. Accretion and Decretion Disks

There are two types of disks that are relevant to the binaries: accretion disks and decretion disks. The circumstellar disks are essentially standard accretion disks. Material flows inward as angular momentum flows outward by means of turbulent viscosity. The angular momentum is carried to the outer edge of the circumstellar disk, where it is removed through tidal torques on the binary.

Decretion disks are less well known. For such disks, the central torque provided by the binary prevents material from accreting. Instead, the binary loses angular momentum to the disk, and the disk then expands outward as the binary contracts (Lin and Papaloizou 1979; Pringle 1991; Lubow and Artymowicz 1996).

A circumbinary disk can in principle behave as a decretion disk. However, initially, during a transient stage, the circumbinary disk can behave as an accretion disk until the disk density profile adjusts to that appropriate for a decretion disk. Another effect is that material might penetrate the gap surrounding the binary and so accrete. Such an effect appears in two-dimensional simulations of disks (Artymowicz and Lubow 1996*b*). In practice, a circumbinary disk might then be intermediate between a pure accretion and a pure decretion disk.

### D. Disk Viscosity Mechanisms

The models for disks described here are based on a simple effective viscosity prescription. The disk viscosity is considered to be an anomalous viscosity resulting from turbulence (see review by Pringle 1981). The disk turbulence could be due to gravitational instability (e.g., Lin and Pringle 1987). The Toomre criterion for gravitational instability in a Keplerian disk can approximately be expressed as $M_d > (H/r)M$, for a disk of mass $M_d$, thickness-to-radius ratio $H/r$, surrounding a binary or single star of mass $M$. Gravitational instabilities are almost certainly important during the earliest phases of star formation, because of the rapid mass buildup of the disk. However, for parameters of some observed systems such as GG Tauri (Dutrey et al. 1994), gravitational instability is also possible.

Another promising mechanism for producing this turbulence is a magnetic shearing instability (Balbus and Hawley 1991). Since this instability relies on the effects of a magnetic field, the exact nature of this instability is complex, and the use of a simple viscosity is a simplification. Dynamos may play a role, as well as various other magnetic instabilities (Tout and Pringle 1992; Hawley et al. 1996).

In application to protostellar disks, this instability depends on the disk being somewhat ionized. It appears possible that protostellar disks are sometimes not sufficiently ionized to be turbulent throughout. Instead, the turbulence may be restricted to the upper layers of the disk that are sufficiently ionized (Gammie 1996; Glassgold et al. 1997).

For the present purposes, these complications will be ignored, and a simple turbulent viscosity will be assumed. The level of disk turbulence is characterized by the Shakura-Sunyaev dimensionless parameter $\alpha = \nu/c_s H$, where $\nu$ is the kinematic turbulent viscosity, $H$ is the disk thickness, and $c_s$ is the disk sound speed (Shakura and Sunyaev 1973).

## III. EFFECTS OF A BINARY ON DISKS

### A. Gravitational Interactions

The tidal forces caused by the binary generally act to distort a disk from its circular form. Since most binaries are eccentric, they bring about time-dependent distortions that occur at the binary frequency and its harmonics. At special locations in the disk that are resonance points, strong interactions between the binary and disk often occur. It is these resonant interactions that usually dominate.

The theory of resonances for disks has been developed as a result of work by Goldreich and Tremaine (1980). From this linear theory, analytic expressions for the torques exerted at resonances can be obtained. To understand how resonances arise in the theory, consider a decomposition of the binary potential $\Phi$ into a sum of rigidly rotating Fourier components:

$$\Phi(r, \theta, t) = \sum_{m,l} \phi_{m,l}(r) \cos(m\theta - l\Omega_b t) \quad (1)$$

where cylindrical coordinates $(r, \theta)$ are centered on the binary center of mass in the inertial frame, and $\Omega_b$ is $2\pi$ divided by the binary orbital period.

For a circular-orbit binary, only diagonal (i.e., $m = l$) terms are nonzero. Setting $m = l$ in equation (1), we see that the potential is static in the frame of the binary, where $\theta - \Omega_b t$ is a constant. In this decomposition, nondiagonal (i.e., $l \neq m$) elements arise only to the extent that the binary is eccentric, which is typically the case. The magnitude of $\phi_{m,l}$ scales with eccentricity as $e^{|m-l|}$. At resonances, waves are launched, which exert torques on disk material as they damp.

Two types of resonances are relevant:

1. Lindblad resonances occur where $\Omega(r) = l\Omega_b/(m \pm 1)$. For $m = l$, their primary effect is to truncate disks. For $m \neq l$, they are called eccentric Lindblad resonances. They can truncate a disk and usually increase the binary eccentricity.
2. Corotational resonances occur where $\Omega(r) = l\Omega_b/m$. They generally damp binary eccentricity. (Technical note: The dynamical effects of corotational resonances depend on the radial derivative of disk vorticity divided by surface density, $\Sigma$. The sign of this quantity might seem ill determined, because the distribution of $\Sigma$ is not known. However, in the case of a disk with a gap, the steep gradient in the density at the disk edge determines the sign of the effects of the corotational resonance. The edges generally produce the dominant corotational torques and act to damp eccentricity.)

## B. Wave Propagation

At Lindblad resonances (LRs), waves are launched that carry energy and angular momentum from the binary. The disk experiences changes in its angular momentum in regions of space where the waves damp. Several mechanisms have been proposed that cause wave damping. The detailed conditions in the disk (which vary over time) determine which mechanism dominates.

Shocks provide one mechanism of wave damping (Spruit 1987; Yuan and Cassen 1994; Savonije et al. 1994). This form of damping may play a role, because a disk is often found in simulations to be truncated near resonances that produce strong waves. Waves that are mildly nonlinear near a resonance might steepen as they propagate away from the resonance and then shock. Such effects are likely to be more important for colder disks. Circumstellar disks that surround protostars may be warm enough that severe damping by this mechanism does not occur. Shocks could provide some accretion over a broad region of a protostellar disk.

Turbulent disk viscosity may act as a dissipation source. However, the level of damping it produces is highly uncertain, especially if magnetic stresses play an important role in providing the turbulence. It is unclear whether the damping of waves due to magnetohydrodynamic (MHD) turbulence can be profitably described through a viscosity. Another complication is that the eddy turnover timescale for the largest turbulent eddies may be longer than the wave period, resulting in a reduced effective disk turbulent viscosity (e.g., Lubow and Ogilvie 1998).

Sound waves can lose energy by radiative damping (Cassen and Woolum 1996). In the absence of radiative damping, the waves would induce adiabatic density fluctuations and propagate without energy loss. When effects of radiative losses are included in the cycle of wave compression and decompression, there is a net loss of energy. The wave radial damping length is roughly equal to the disk thickness times the ratio of the disk cooling time to the local orbit period in the disk. The range of plausible protostellar disk conditions permits rapid or slow decay to occur by this process.

Nearly all studies of waves in disks have regarded the disk as two-dimensional. That is, dynamical effects in the vertical direction are often ignored or simplified. In many circumstances, disks are likely to be optically thick in the vertical direction, and substantial vertical temperature variations could occur. For these conditions, conventional wisdom was that sound waves ($p$ modes) would be launched at a Lindblad resonance, just as in the vertically isothermal case. The waves would then experience a propagation speed that varies with height. As a result, the waves would refract up into the disk atmosphere, where they would shock, over a radial distance scale of the order of the disk thickness (e.g., Lin et al. 1990).

However, recent semianalytic 3D results provide a different picture. For a disk having a midplane temperature that is large compared to the disk surface temperature, $f$-mode waves are launched by Lindblad resonances (Lubow and Ogilvie 1998). The waves are naturally confined vertically by the increasing vertical gravity with height from the disk plane and do not propagate vertically (the wave is vertically evanescent). As the $f$ mode propagates radially away from the resonance, it behaves like a surface gravity wave. The wave energy becomes increasingly confined (channeled) close to the disk surface. The wave amplitude grows, and wave damping by shocks can sometimes occur (Ogilvie and Lubow 1999).

As a consequence of the damping mechanisms described above, it appears plausible that the waves launched from resonances are damped locally, somewhat near the resonance in protostellar circumbinary disks. Nonlinear simulations provide some support for local damping (e.g., Artymowicz and Lubow 1994). (However, numerical artifacts may be playing a role.) We will often assume local damping in the discussion below.

### C.  Gap Formation

Where wave damping occurs, the angular momentum of the disk material changes. This process leads to the opening of a gap in the disk (Lin and Papaloizou 1979; Goldreich and Tremaine 1980). Gap clearing is observed in astrophysical disks, among others in planetary rings perturbed by satellites (e.g., Lissauer et al. 1981). Viscous effects in the disk act to fill the gaps. The criterion for gap opening at any radius in a viscous disk is that the viscous torque (which acts to smooth density variations) must be less than the resonant torque. Both torques are typically evaluated using linear theory. In the case of an eccentric binary with moderate $e$, many resonances are present, but only one or two dominate. The gap opening criterion has been found to agree with the results of SPH simulations of eccentric binaries (Artymowicz and Lubow 1994).

If one assumes that the resonant torque is exerted locally near the resonance, the disk edge location is determined by the location of the weakest resonance for which the resonant torque equals or exceeds the viscous torque. If waves are damped nonlocally, then the viscosity-balancing resonance is still the location where the disk is effectively truncated, although the overall maximum density of the disk may be found farther away from the perturber (Takeuchi et al. 1996). In the extreme case of negligible viscosity and little other wave damping, there would only be a minimal gap (due to orbit crossing). In any case, consideration of the optical depth of the disk is needed to predict the position of the edge observed in a particular wavelength band. As a rule, the disks are sufficiently opaque to make the torque-balance definition of the gap size useful.

Figure 2 shows the expected locations of the inner edge of a typical CB disk for various values of binary eccentricity and disk turbulent viscosity. The dominant resonance responsible for disk truncation in a binary system having $e \sim 0.2$ and $\alpha \sim 0.01$ is typically the $m = 2$, $l = 1$ outer

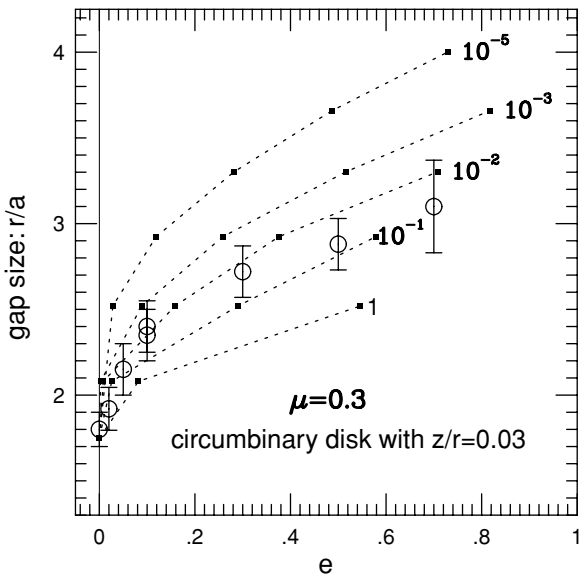

Figure 2. The size of a gap around a binary system with eccentricity $e$, in terms of binary semimajor axis $a$, for different $\alpha$ parameters for viscosity of the disk (numbers give $\log \alpha$).

Lindblad resonance (LR). All the barlike, $m = 2$ gravitational field harmonics decrease slowly with distance, in comparison with $m > 2$ harmonics. The noneccentric $(m, l) = (2, 2)$ LR is so strong that it easily clears the gas from its vicinity.

The importance of the next-strongest (2,1) LR is connected with the fact that binaries with $e = 0.2$–$0.5$ spend relatively much time near apastron, where their angular speed decreases to $\Omega_b/(1 + e)^2 \sim \Omega_b/2$. This gives rise to a strong $l = 1$ harmonic, which rotates uniformly at the speed $l\Omega_b/m = \Omega_b/2$. Whenever that resonance is too strong compared with viscosity, as happens at intermediate and high eccentricity $e$ of the binary, the disk recedes further to the regions of higher-$m$ LRs with $l = 1$.

Plots similar to Fig. 2, showing the expected disk sizes for circumprimary and circumsecondary disks (Artymowicz and Lubow 1994), demonstrate that as the eccentricity increases, the $m = 2$ inner eccentric LRs with increasing $l$ number ($l > 2$) play a dominant role in CS disk truncation.

Jensen et al. (1996) and Jensen and Mathieu (1997) have successfully fit the infrared spectral energy distributions of pre-main-sequence (PMS) binaries with a simple model of disks with gaps. The standard protostellar disks are sharply truncated at radii corresponding to the torque balance. Exceptional cases, such as AK Scorpii, require, however, that the gap be partially filled with dust, probably embedded in gas flowing through the gap from a CB disk (see the chapter by Mathieu et al., this volume).

### D. Generation of Disk Eccentricity and Features

A disk can become eccentric as a result of its interaction with the binary. The eccentricity can arise as an instability in a system with a binary in circular orbit or through direct driving by an eccentric binary.

One example of an eccentric instability is that which occurs in CS disks of superhump binaries (Whitehurst 1988). The instability occurs through a process of mode coupling and relies on the effects of the 3:1 resonance in the disk (Lubow 1991). For this instability to operate, the binary mass ratio must be fairly extreme, greater than 5:1. On the other hand, it cannot be too extreme, or the instability will be damped by viscosity. This or other mode-coupling instabilities in principle could also operate in a CB disk at the 1:3 resonance [see equation (35) of Lubow 1991, with $m = 1$].

An eccentric binary with unequal stellar masses drives eccentricity in an initially nearly axisymmetric CB disk via its one-armed bar potential with $(m, l) = (1, 0)$. The disk disturbance follows the slow apsidal motion of the binary (Artymowicz and Lubow 1996a). The one-sidedness of disk forcing has a twofold effect. Initially the eccentricity of the disk's edge, denoted as $e_d$, grows as

$$\dot{e}_d = -\frac{15}{16} e\mu(1 - \mu)(1 - 2\mu)(a/a_d)^3(1 - e_d^2)^{-2} \sin\varpi \quad (2)$$

where $a$ and $e$ are the binary's semimajor axis and eccentricity, $a_d$ is the semimajor axis of the disk inner edge, and $\mu$ is the binary mass parameter (secondary mass divided by total mass). The longitude of periastron of disk orbits, with respect to the stellar periastron, is denoted as $\varpi$. Typically, disk eccentricity grows significantly in a few hundred orbital periods $P$ of the binary, while $a_d \approx$ constant. During this phase, $\varpi$ is locked at a stable value $\varpi = 3\pi/2$, corresponding to the perpendicular relative orientation of the two (disk and binary) apsidal lines. However, as $e_d$ grows, the locking action of the lopsided potential weakens, and the standard prograde precession of the disk (due to the quadrupole moment of the double star) dominates. Both the analytical and SPH results show that the disk edge begins to precess around the binary when $e_d/e \sim 0.2$–$0.7$. Afterwards, eccentricity oscillates with a precessional period ($\sim 10^2$–$10^3 P$), driven up and down by the $\sin\varpi$ factor in equation (3). In numerical calculations, the CB disk typically attains $e_d/e = 0.5$–$1$. In addition to the direct driving described by equation (2), eccentric instabilities may play a role. Mode-coupling instabilities appear to saturate in the end state of the SPH simulations, which, apart from the periodic driving, resembles the free $m = 1$ mode studied by Hirose and Osaki (1993). In this mode, pressure gradients between adjacent rings synchronize their precession, counteracting the differential quadrupole-induced precession.

Figure 3 presents the eccentric disk evolution in snapshots from a very long SPH simulation of a CB disk around a binary with $e = 0.8$ and $\mu = 0.3$. The upper left panel of the figure illustrates the initial growth of $e$, while the remaining three panels show a quasistationary (oscillating) eccentricity of the dense, precessing, disk edge. The binary itself is always shown at periastron and, because of its very high eccentricity, is not well resolved in this figure.

Streamlike features emanating from the disk edge toward the otherwise empty circumbinary gap are nonresonant in origin. They are seen only on the receding part of the gas trajectories, following the periastron, when the perturbation by the binary is greatest. Since the binary is revolving much faster than the disk material, its main ($m = 1$) harmonics create disturbances that are later enhanced by the kinematics of orbits crossing at caustics. These disturbances lead to local shocking of gas along the feathery features that are colliding with the disk rim. The shocks occur before the material starts descending toward the binary along the elliptic outline of the disk edge (hence the asymmetry between the two halves of the elliptic edge). The cuspy features at the disk rim are potentially observable. Their number contains information about the mean gap size, $a_d/a$, and the mean binary separation $a$ (which may not be easy to establish otherwise from observations of wide binaries). They also act as potential sites for the streams of gas flowing onto the binary, described later.

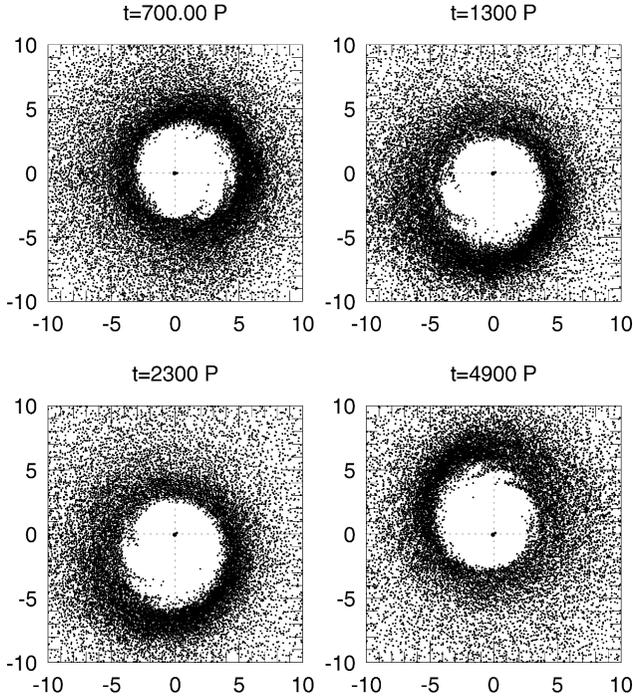

Figure 3. Top views of an SPH simulation of an eccentric binary ($e = 0.8$) with mass ratio 3:7. Axes are scaled in units of $a$, and time is given in binary orbital periods $P$. Slow precession and evolution of disk eccentricity are apparent, as well as the transient features generated at the disk edge by the binary.

### E. Noncoplanar Disks

The binary tidal field might be expected to force a disk to coplanarity because of differential precession. Differential precession occurs because the disk nodal precession rate varies with radius. Consider a disk as a collection of ballistic orbits that initially all have the same tilt and line of nodes, so the disk is of a tilted planar form. Over time, different annuli in the disk would develop different nodal phasing, and the disk coherence would be lost. For a fluid disk, one might expect that the strong nonplanar shearing motions would give rise to strong dissipation (Papaloizou and Pringle 1983), which would lead to coplanarity with the orbit plane of the binary.

For a fluid disk, there is also the possibility that the disk could act as a coherently precessing body. Theory (Papaloizou and Terquem 1995) and simulations (Larwood et al. 1996) suggest that this is indeed possible, provided that the radial sound crossing time in the disk is shorter than the timescale for differential precession. For conditions in binaries, coherent precession typically requires that the ratio of the disk thickness to radius be greater than about 0.05. When the ratio lies in the range of 0.05 to 0.1, mildly warped structures develop. For much thinner disks, with thickness ratio of 0.02 or less, there is a strong disruption of the disk structure by differential precession.

The origin of the nonplanarity is unclear. It may be established by star formation in widely separated pairs. Observations of visual binaries suggest that nonalignment of the stellar spin axes occurs for separations greater than about 40 AU (Hale 1994). The stellar spin axis direction may reflect the orientation of a preexisting surrounding disk, which was subsequently accreted. Consequently, widely separated binaries might well have formed with misaligned disks.

A recent HST image of HK Tau provides evidence for a misaligned, tidally truncated disk in a binary system (Stapelfeldt et al. 1998). Although the disk's orientation is clear from the image (nearly edge-on), the binary's orientation is unknown. From statistical arguments concerning the binary orbit, the observed disk size suggests that the disk is tidally truncated. With moderate binary eccentricity ($e < 0.5$), the degree of misalignment is likely to be at least 20 degrees.

For this system, the disk thickness-to-radius ratio appears to be in the range for coherent precession with warping. Furthermore, the disk precession period is plausibly of the order of the age of the binary (several million years). The expected warping might then be related to the observed asymmetries.

For young stars of high luminosity (mass exceeding a few $M_\odot$, $L$ greater than $\sim 10 \, L_\odot$) the radiation from the central star can induce a warp instability in a CS disk (Armitage and Pringle 1997). The warp is strongest in the inner parts of the disk but may become significant in the outer parts.

**F. Mass Flow through Gaps**

Until the mid-1990s, the consensus, based in part on numerical case studies, was that there is essentially no flow of gas from the CB disk onto the central binary. Therefore the evolution of CS disks was thought to be independent of the CB disk. One of the natural consequences would be a faster depletion of the CS disks in binaries, because of their shorter viscous evolution time. However, the evidence from the accretion rates onto PMS spectroscopic binary stars (presumably from the CS disks) was in disagreement with naive expectations (Mathieu 1994).

It now appears that stars in single and binary systems may accrete similar amounts of mass from surrounding disks. Two-dimensional simulations of disks in binary-disk systems with moderately thin and viscous disks have revealed the presence of a flow through the gap from the CB disk, generally in the form of two gas streams (Artymowicz and Lubow 1996*a,b*). One reason the old paradigm was inadequate was its reliance on a one-dimensional approximation for the physical quantities. The mass density, resonant torque, and viscous torque were treated as functions of radius only, by considering their azimuthal averages. The two-dimensional studies showed that at certain locations in azimuth, the conditions may be such as to permit mass to penetrate the gap.

The flow process for moderately eccentric binaries can be understood in terms of an effective potential. The inner edge of the CB disk is maintained by the outer Lindblad resonance of the "simplified potential" or the combined $(m, l) = (0, 0)$ and $(2, 1)$ potential (see Artymowicz and Lubow 1996*a* for comparative simulations with the full and the simplified potentials). This bisymmetric potential is static in a frame that rotates at the rate $\Omega_b/2$, or one-half the mean angular speed of the binary. Its effective potential (gravitational potential plus centrifugal barrier) has two saddle points, which are unstable to small perturbations. They correspond closely to the collinear Lagrange points L2 and L3 in the circular restricted three-body problem of celestial mechanics. Free particles that have a small inward velocity would flow inward as a stream from the saddle points, similar to what occurs in a classical Roche overflow process. The finite enthalpy of the disk gas provides an expanding flow that produces the two streams, which are directed toward the stars. We have proposed that an "efficient" flow of gas from the CB disk can sometimes occur. (An efficient flow is one that produces an inward mass flow rate that is comparable to the rate that would be produced in the absence of the binary.)

The CB disk is typically truncated just outside the (2,1) LR by the resonant torques it produces. In this case, the gas passes through the vicinity of points at which penetration inward is easiest (i.e., the saddle points corotating with the simplified potential), located inside the LR. An efficient flow requires (and results from) the Bernoulli constant of the gas at the disk edge being sufficient to overcome the effective potential barrier between the LR and corotation point. This condition can be satisfied in disks which are sufficiently warm (larger enthalpy) or viscous (disk edge closer to the center and endowed with larger effective energy). We have found such flows in SPH simulations of disks with relative thickness ratio $H/r \approx c_s/\Omega r \geq 0.05$ and $\nu \sim 10^{-4}\Omega r^2$.

Figure 4 shows the snapshots from one SPH simulation, modeling a binary system with nearly equal masses (mass parameter $\mu = 0.44$) and eccentricity $e = 0.5$. The crosses denote the saddle points of the $(0, 0) + (2, 1)$ potential, and the added line segments indicate the preferred direction along which gas flows through these points. The direction is determined by taking into account the underlying potential and the Coriolis force, but neglecting gas enthalpy and velocity before the passage (cf. Lubow and Shu 1975). We see that the simplified binary potential indeed determines the number of streams, their rotation at one-half the mean binary rate, and the pitch angle of the streamlike features.

The mismatch of the streams' angular velocity with that of the binary causes a marked time dependence of the mass accretion rate onto the stars (or their CS disks). Although the binary pulls on the disk most strongly at apastron, the material does not impact the binary in Fig. 4 until approximately periastron. The profiles of $\dot{M}(t)$ depend sensitively on the binary's orbital elements, making the variability a valuable tool in analyzing the properties of unresolved PMS systems (Artymowicz and Lubow 1996*b*).

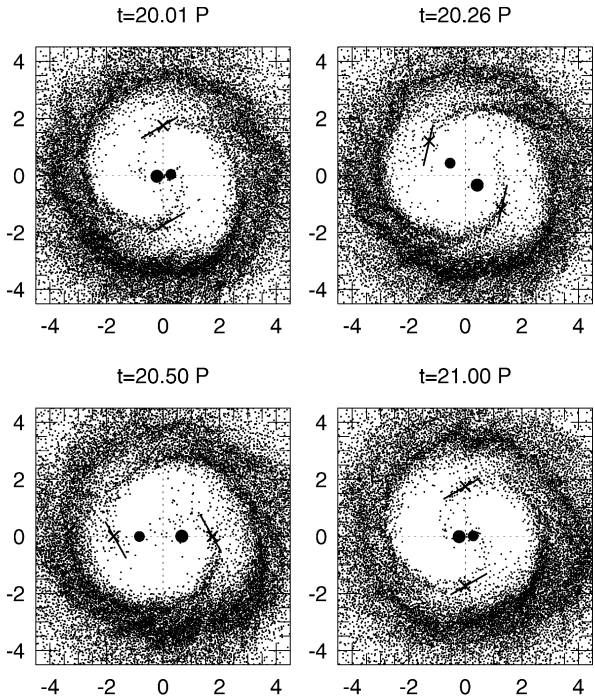

Figure 4. Circumbinary disk with half-thickness $H/r \approx c_s/\Omega r = 0.05$ transferring mass onto the central eccentric ($e = 0.5$) binary with nearly equal components (mass ratio 11:14).

Near the stars, in the case of moderate orbital eccentricity, the classical Roche effective potential (which corotates with the mean motion of the binary) exerts control over the flow. The flow is channeled by this potential in the vicinity of the classical outer Lagrange points L2 and L3. The L3 point (on the side of the less massive component) allows easier penetration because its effective potential barrier is lower. The L3 point therefore produces a higher accretion rate than the L2 point [as found by Artymowicz (1983) and by Bate and Bonnell (1997)]. Two major consequences of this flow are (i) accretion rate reversal, which may cause the less massive star to become more luminous (due to a higher accretional luminosity), and (ii) mass equalization. The importance of these effects for interpretation of observations is not yet clear (Clarke 1996). In principle, there are implications for the distribution of mass ratios and the observed binary frequency among all types of stars, because actively accreting stars with small companions may be easier to detect as binaries.

Several observations, summarized below, support the existence of a mass flow through gaps in CB disks.

Near-IR images of classical T Tauri binaries GG Tau and UY Aur (Roddier et al. 1996; Close et al. 1998) revealed the inner edges of the CB disks. The inner edge locations based on millimeter observations (Dutrey et al. 1994; Duvert et al. 1998) agree with the IR imaging and theoretical expectations, and the disk rotation is consistent with Keplerian. The IR observations provide weak evidence for the presence of material in the central gap. A high-velocity feature may be associated with gas streams in UY Aur.

A double-lined cTTS close binary, DQ Tau, does not have much space for CS disks, yet it shows all the signs of active accretion, presumably through the CB gap. Continuum and line fluxes undergo modulations on the period of the binary (Mathieu et al. 1997; Basri et al. 1997). The highest fluxes occur near binary periastron, a behavior consistent with the SPH simulations, if the enhanced emission is due to the gas stream impact onto the stars or very small CS disks. The dynamical model is consistent with the variability of spectral features, which provides an independent test for the model.

Spectral energy distributions of several PMS binaries (Jensen and Mathieu 1997) show no evidence for a binary-produced gap. This may indicate that the dust emission from the material flowing through the gap masks its presence.

Observations of 144 G and K dwarfs in Pleiades (Bouvier et al. 1997) showed no significant differences in the rotational velocity distributions for single and binary stars. This suggests that accretion onto both types of stars is similar and, in the case of binaries, likely proceeds from a CB disk.

Embedded IR companions to T Tauri stars (Koresko et al. 1997; Meyer et al. 1997) are plausibly interpreted in terms of the mass flow from CB disks onto the secondary companions. Cases in which the more massive star is apparently accreting more gas are common (Monin et al. 1998).

## IV. EFFECTS OF DISKS ON A BINARY

The gravitational interaction of the bulk of the disk with the binary changes its orbital elements. Gas streams can also change the orbital elements, both by gravitational torque contributions and by direct impact (advection of mass, energy, and angular momentum). The potentially important effects of gas streams depend on the still unknown details of where and with what velocity the streams hit the CS disks or stars, in the highly dynamic environment of an eccentric binary. (However, in section V we mention preliminary work on protoplanets, treated as binaries with low eccentricities and extreme mass ratios.) In this section we describe the long-range effects of the disks, which do not include effects of gas streams.

## A. Binary Separation and Eccentricity

In a binary with mass parameter $\mu = 0.3$ and eccentricity $e = 0.1$, studied by Artymowicz et al. (1991), by far most of the semimajor axis and eccentricity driving is due to an outer LR of the $(m, l) = (2, 1)$ potential component (which also keeps the disk edge in equilibrium). In such a case, the gravitational torque $T_{ml}$ is equal to the axisymmetric viscous torque, $T_{21} = 3\pi\Sigma\nu\Omega r^2$, evaluated at radius $r$ in the disk just outside of the edge region (maximum wave damping region). This relationship provides a means of estimating the rate of decrease of the binary semimajor axis $a$ (cf. Lubow and Artymowicz 1996):

$$\frac{\dot{a}}{a} = -\frac{6l}{m}\frac{\alpha(H/r)^2}{\sqrt{1-e^2}\mu(1-\mu)}\frac{a}{r}q_d\Omega_b \tag{3}$$

where $q_d = \pi r^2 \Sigma/M$ quantifies the disk-binary mass ratio, and $M$ is the total mass of the binary. This robust formula is consistent with the numerical results reported by Artymowicz et al. (1991), giving $\dot{a}_b/a \sim -10^{-3} q_d \Omega_b$ for the binary with $\mu = 0.3$ and eccentricity $e = 0.1$.

The eccentricity evolution of a binary is dominated by the effects of the CB disk. In general, eccentricity grows at a rate

$$\dot{e} = \frac{1-e^2}{e}\left(\frac{l}{m} - \frac{1}{\sqrt{1-e^2}}\right)\frac{\dot{a}}{a} \tag{4}$$

This equation implies that a CB disk shepherded by the (2,1) potential in a low eccentricity binary has $e\dot{e} \simeq -\dot{a}/(2a)$, in good agreement with the SPH results of Artymowicz et al. (1991).

The numerical ratio between rates $\dot{e}/e$ and $\dot{a}/a$ was analyzed by Lubow and Artymowicz (1996), who concluded that the ratio of these rates increases linearly with $e$ up to a maximum found at $e \sim 0.03$ (for their assumed disk parameters), then decreases as $1/e$. The maximum driving corresponds to a short timescale for doubling eccentricity, of the order of several hundred binary orbit periods.

At higher values of eccentricity, many resonances are excited, including resonances that damp eccentricity (eccentric inner LRs and corotational resonances). At an eccentricity of about 0.5 to 0.7, these resonances nearly cancel, and the mean rate of growth of eccentricity diminishes considerably (Lubow and Artymowicz 1993).

In addition, in systems with eccentric disks, the disk forces the binary eccentricity to oscillate on the relative precessional timescale. This occurs because of the disk's (1,0) one-sided forcing, which is similar in basic physics to the reverse influence of the eccentric binary on the disk, discussed previously. It is not uncommon to find in SPH simulations that this leads to temporary (nonsecular) eccentricity damping periods in many binary-disk systems.

## B. Implications for Planet Formation

A paradigm in the planet formation theory was that a planet (specifically Jupiter) stops accreting gas when it opens a gap in the primordial nebula (e.g., Lin and Papaloizou 1993). Artymowicz and Lubow (1996*b*) suggested that this does not normally happen (cf. Artymowicz 1998 for a sample SPH calculation). Instead, simulations suggest that matter flows through an otherwise nearly evacuated gap, much as in the case of a binary star (although the dynamics may be different). This may provide a method of formation in standard solar nebulae for the "superplanets," or planets with masses $\sim 5$ $M_{Jup}$ (Jupiter masses) or greater. Newly discovered extrasolar planets around 14 Herculis, 70 Virginis, Gl 876, and HD 114762 may fall in this mass range. This somewhat arbitrary mass definition for superplanets assumes a solar-mass companion. In the theory, the mass ratio, rather than the mass, is relevant. Consequently, the dynamical properties we ascribe to superplanets may be appropriate to 70 Vir, since its minimum mass ratio is in the range for superplanets. Mutually compatible results supporting this general scenario have since been obtained with five modern hydrodynamical schemes in eight implementations (Artymowicz et al. 1999; Kley 1999; Bryden et al. 1999; Lubow et al. 1999; chapter by Lin et al., this volume). Results show that there is no fundamental impediment in the growth process of a giant protoplanet via disk accretion to become a superplanet or possibly even a brown dwarf-mass body (although not a brown dwarf according to formative and structural definition). The difficulties are more "practical": The mass of the gas in the disk or its longevity may be insufficient, or the disk may have too little viscosity or pressure to supply efficient flow. For instance, relatively thin and low-viscosity disks tend to prevent the growth of superplanets (see the chapter by Lin et al., this volume). Sufficiently high-mass perturbers may limit further accretion by tidal effects. For standard moderate-mass nebulae (with mass several times the minimum solar nebula, viscosity parameter $\alpha \sim 10^{-3}$–$10^{-2}$, and $H/r \approx 0.05$) one finds that Jupiter could easily double its mass within a few Myr. Present masses of the solar system giant planets may thus indicate that they formed late in the life of the solar nebula and could not capture the dwindling supply of mass. An apparent lack of their large-scale inward migration supports speaks of the same. That may not be a universal outcome, however.

Color Plate 12 demonstrates the flow, simulated with the PPM (piecewise parabolic method) for a disk with $\alpha = 0.004$ and a Jupiter-mass protoplanet in circular orbit. A top view of a fragment of the 2D disk with embedded Jupiter-mass planet is shown. Two wakes, which are shock waves, penetrate the gap. Disk gas hitting restricted sections of the wakes close to the Lagrange points delimiting the Roche lobe (white oval) is brought to a near-standstill in the frame corotating with the planet and falls toward the planet.

An intriguing correlation between the mass (ratio) of the planet and its orbital eccentricity exists among the currently confirmed radial-velocity planetary candidates (Mazeh et al. 1997; Marcy and Butler 1998). We reproduce in Fig. 5 the distribution of extrasolar planets and solar system planets, including the pulsar planets, in the $m$-$e$ plane. This distribution, unless an artifact caused by small statistics, shows that superplanets avoid circular orbits and that the majority of Jupiter-class and smaller planets prefer circularity.

In the case of planets, the gaps are much smaller than in the case of binary stars, and the tidally disturbed material experiences many more resonances from the higher-$m$ components of the planet's potential. Still, at some critical mass (called the crossover mass) the (2,1) potential harmonic dominates (the gap must extend to at least the 1:2 and 2:1 commensurabilities), and the same mechanism as described for stars may rapidly deform the initially nearly circular (super)planet orbits into ellipses of intermediate $e$ ($e > 0.2$). On the other hand, planets unable to open gaps, typically less massive than Neptune, suffer strong and model-independent eccentricity damping by coorbital Lindblad resonances (Ward 1986; Artymowicz 1993). In addition, the Jupiter-class planets that open a modest gap are also circularized by corotational torques (Goldreich and Tremaine 1980). Thus, although the value of the crossover mass (which likely depends on disk temperature and viscosity) has yet to be determined in high-resolution calculations taking into account mass flows, its existence is well established. Based on early SPH calculations of only the gravitational interactions, the crossover mass in a standard solar nebula was determined to be $\sim 10$ $M_{Jup}$ (Artymowicz 1992). In more recent work, we found evidence that gas inflows such as seen in Color Plate 12 generally damp eccentricities. Since the flow is inefficient at masses exceeding $\sim 10$ Jupiter masses (in a standard solar nebula), the crossover mass may indeed be close to that value. However, given the diversity of nebular properties, we do not expect a unique $e(m)$ functional dependence. Rather, a pure disk-planet interaction in an ensemble of systems would result in a switch from very low to moderately high eccentricities over a finite range of masses, probably centered near $m = 10$ $M_{Jup}$. Should such a mass-eccentricity pattern be confirmed by future, improved statistics, it would strongly argue for the importance of binary-disk interactions in shaping the orbits of exoplanets. Conversely, a large percentage of Jupiter-like or smaller planetary candidates found on high-$e$ orbits around single stars would require a complementary mechanism for eccentricity generation: planet-planet interaction (e.g., Weidenschilling and Marzari 1996; Lin and Ida 1997; Levison et al. 1998).

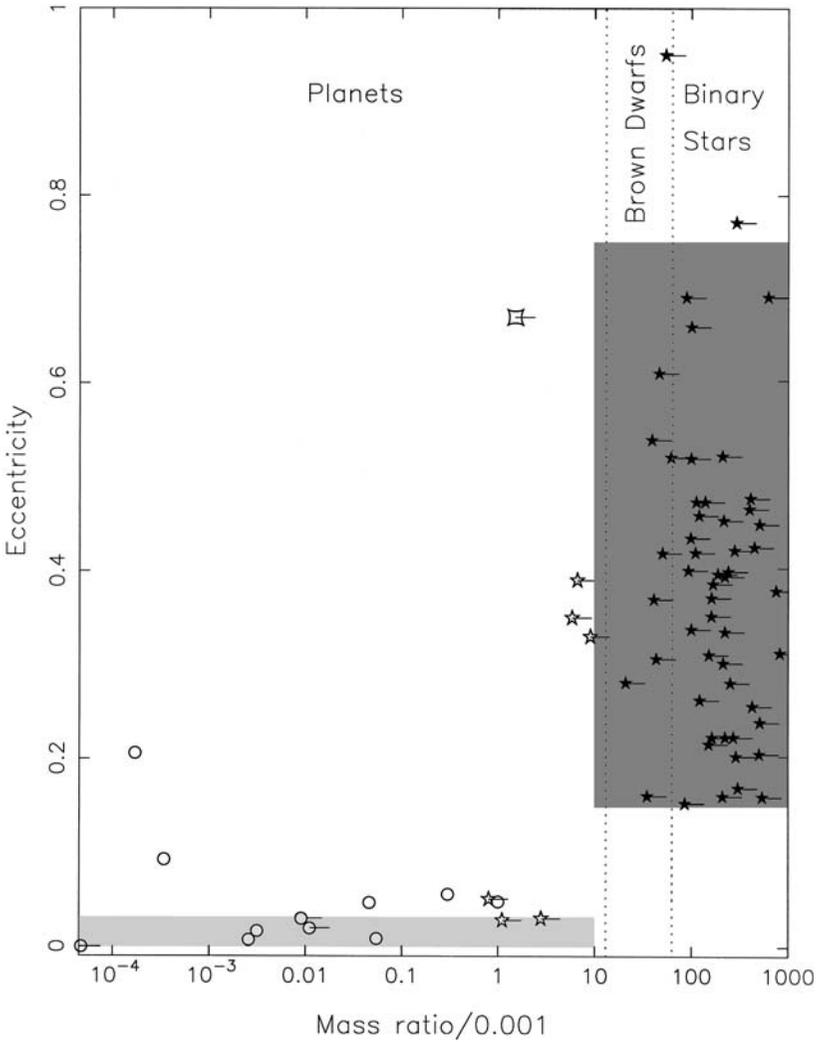

Figure 5. The *m-e* correlation vs. the predictions of the disk-planet interaction theory, in which the eccentricity is damped by protoplanetary disks at mass smaller than approximately 10 Jupiter masses and rapidly grows at larger masses. Potentially tidally circularized companions have been omitted (cf. Artymowicz 1998; Artymowicz et al. 1999).

Independent evidence of disk-planet interaction in extrasolar systems, resulting in migration, is provided by the short-period "hot" Jupiters (Lin et al. 1996; chapter by Lin et al., this volume).

## V. SUMMARY

Disks play an important role in the evolution of young binaries. At the earliest phases of evolution, most of the material accreted by a sufficiently close binary (semimajor axis less than about 20 AU) passes through a circumbinary disk (see Fig. 1). The evolution occurs in the dynamical collapse timescale of about $10^5$ years. During the later T Tauri phase, the evolution occurs on a viscous timescale of the disk. The binary undergoes eccentricity excitation because of its interactions with the disk.

Many aspects of disk-binary interactions are directly observable and provide a potentially powerful diagnostic tool. Resonances emit spiral waves and clear gaps (Fig. 2). Simulations and analytic theory indicate that circumbinary disks sometimes become eccentric as a result of their interactions with the central binaries (Fig. 3). Small-scale features at the disk edge result from interactions with an eccentric binary.

Mass sometimes flows as gas streams from the circumbinary disk, through the central gap, onto the binary (see Fig. 4). This flow can cause the secondary star to appear to be the more luminous (luminosity reversal) and can cause the secondary to accrete more mass than the primary (tending toward mass equalization). The flow has additional consequences on the orbital evolution of the binary that remain to be determined. There are several observational indications of the existence of these flows.

The analysis of binary-disk interactions can be extended to the case of planet-disk interactions. Mass flow onto a young Jupiter can occur despite the presence of a gap (Color Plate 12). It appears possible that some planets can grow to mass ratios with their stars comparable with or larger than the minimum mass ratios of the newly discovered extrasolar superplanets, such as those in 70 Vir or 14 Her. Unlike planet-planet interaction, planet-disk interactions may produce substantial eccentricity for high-mass planets and low eccentricity for low-mass planets. Current observations provide some support for this correlation, but the sample size is small (Fig. 5). The pattern of population of the mass-eccentricity plane by exoplanets should be indicative of which process is typically responsible for the shape of their orbits.

*Acknowledgments* This work was supported by NASA Grants NAGW-4156 and NAG5-4310. S. L. acknowledges support from The Isaac Newton Institute for Mathematical Sciences of Cambridge University. P. A. acknowledges the STScI visitor program and Swedish Natural Science Research Council research grants. We thank the referee, Matthew Bate, for comments.